\documentclass[conference]{IEEEtran}
\IEEEoverridecommandlockouts
\usepackage[normalem]{ulem}
\usepackage{cite}
\usepackage[most]{tcolorbox}
\usepackage{amsmath,amssymb,amsfonts}
\usepackage{algorithmic}
\usepackage{graphicx}
\usepackage{textcomp}
\usepackage{multirow}
\usepackage{xcolor}
\usepackage{hyperref}
\usepackage{listings}
\usepackage[hyphens]{xurl}
\usepackage{fancyvrb}
 \usepackage{booktabs} 
 \usepackage{hyperref}
\usepackage{cleveref}
\usepackage{soul}
\usepackage{etoolbox}

\lstdefinelanguage{dafny}{
  morekeywords={
    class, datatype,codatatype, type, iterator, lemma,
    bool, nat, int, object, set, multiset,seq, array,array2,array3, map,
    function, predicate,copredicate,
    pthost, var, static, refines,
    method, constructor,comethod,
    return, returns, yields, abstract, module, import, default, opened, as, in,
    requires, modifies, ensures, reads, decreases, free,
    match, case, false, true, null, old, fresh, choose, this,
    assert, assume, print, new, if, then, else, while, invariant, break, label,    return, yield, parallel, where, calc
  },
  sensitive=true,
  morecomment=[l]{//},
  morecomment=[s]{/*}{*/},
  morestring=[b]"
}

\lstdefinelanguage{Dafny}{
    keywords={method, function, predicate, class, constructor, trait, extends, return, returns, requires, ensures, modifies, reads, invariant, forall, exists, if, else, true, false, while, for, match, case, var, const, int, nat, real, bool, bv4, char, string, array, set, multiset, seq, map, new, print},
  keywordstyle=\color{blue}, %
    identifierstyle=\color{black},
    comment=[l]{//},
    commentstyle=\color{gray}\ttfamily,
    stringstyle=\color{red}\ttfamily,
    sensitive=true
}

\definecolor{mygreen}{HTML}{117733}
\definecolor{mygreen}{HTML}{117733}
\definecolor{myred}{HTML}{8A1414}
\definecolor{grey}{RGB}{89, 89, 89}
\definecolor{blue}{RGB}{38, 105, 202} 
\definecolor{red}{RGB}{204, 102, 119} 
\definecolor{turquoise}{RGB}{68, 170, 153}

\lstdefinestyle{dafnystyle}{
 basicstyle       = \ttfamily\footnotesize,
  language        = Dafny,
  keywordstyle    = \color{mygreen}\bfseries,
  commentstyle    = \color{gray}\itshape,
  stringstyle     = \color{red},
  frame           = tb,
  numbers         = left,
  numberstyle     = \tiny\color{gray},
  breaklines      = true,
  showspaces      = false,
  showstringspaces= false,
  emph={Passing, Failing, Main, demo},
  emphstyle=\color{myred}
}

\def\BibTeX{{\rm B\kern-.05em{\sc i\kern-.025em b}\kern-.08em
    T\kern-.1667em\lower.7ex\hbox{E}\kern-.125emX}}

\definecolor{grey}{RGB}{89, 89, 89}
\definecolor{blue}{RGB}{38, 105, 202} %
\definecolor{red}{RGB}{204, 102, 119} %
\definecolor{turquoise}{RGB}{68, 170, 153} %
\tcbset{
  myboxstyle/.style={
    colback=turquoise!5,
    colframe=turquoise,
    colbacktitle=turquoise,
    boxrule=0.4pt,
    arc=1mm,
    left=6pt,
    right=6pt,
    top=0pt,
    bottom=0pt,
    fonttitle=\bfseries,
    width=\columnwidth,
  }
}

\begin{document}

\title{Improving Debugging in Verification-Aware Languages Through Automated Fault Localization: A Case Study in Dafny \\
\thanks{\IEEEauthorrefmark{1} Author order follows a senior-author-last convention.}}

\author{
\IEEEauthorblockN{1\textsuperscript{st} Álvaro Silva}
\IEEEauthorblockA{
\textit{INESC TEC, Faculdade} \\
\textit{de Engenharia,} \\
\textit{Universidade do Porto}\\
Porto, Portugal \\
amfpsilva@hotmail.com %
}
\and
\IEEEauthorblockN{2\textsuperscript{nd} Isabel Amaral}
\IEEEauthorblockA{
\textit{INESC TEC, Faculdade} \\
\textit{de Engenharia,} \\
\textit{Universidade do Porto}\\
Porto, Portugal \\
isabel.andre.amaral@gmail.com 
}
\and
\IEEEauthorblockN{3\textsuperscript{rd} João Pascoal Faria}
\IEEEauthorblockA{
\textit{INESC TEC, Faculdade} \\
\textit{de Engenharia,} \\
\textit{Universidade do Porto}\\
Porto, Portugal \\
jpf@fe.up.pt %
}
\and
\IEEEauthorblockN{4\textsuperscript{th}Alexandra Mendes\IEEEauthorrefmark{1}}
\IEEEauthorblockA{
\textit{INESC TEC, Faculdade} \\
\textit{de Engenharia,} \\
\textit{Universidade do Porto}\\
Porto, Portugal \\
alexandra@archimendes.com %
}

}

\maketitle

\begin{abstract}
Verification-aware languages, like Dafny, integrate formal specifications directly into source code to enable static correctness checks. However, when verification fails, the feedback provided is often limited to the specific condition of the error, such as a violated postcondition, rather than the root cause of the fault. For example, in Dafny, while its counterexample features provide concrete execution traces, they typically expose a single failing path per assertion failure, leaving the developer to manually inspect the entire trace to locate the error.

This paper investigates automated fault localization for verification-aware languages by comparing two paradigms: state-based and counterexample-based localization. Our \emph{state-based} localization strategy replicates the ``snapshot'' methodology of AutoFix by inferring invariants and predicates to identify suspicious program states. The \emph{counterexample-based} strategy consists of a family of techniques that progressively enrich the use of verifier output: from raw counterexample extraction, to structured single-trace ranking, and to multi-trace aggregation.
 
To validate these methods, we present an evaluation framework that uses MutDafny to generate a diverse mutant dataset from DafnyBench and measures localization effectiveness using the EXAM score. Our results show that counterexample-based approaches substantially outperform state-based localization in this setting. Structured ranking over a single trace yields the largest improvement over raw counterexample output, while multi-trace aggregation provides additional gains in robustness and debugging utility by increasing coverage and reducing path bias introduced by the solver. These findings demonstrate that effective fault localization in verification-aware languages depends both on using counterexample information and on how that information is structured and diversified.
\end{abstract}

\begin{IEEEkeywords}
Fault localization, verification-aware languages, Dafny, counterexamples, design by contract
\end{IEEEkeywords}

\section{Introduction}
\label{sec:introduction}

Software is embedded in all aspects of modern life, from smartphones to medical devices, making software correctness more critical than ever. Yet, writing software free of bugs remains challenging, and locating the root cause of a bug is often an arduous and time-consuming part of debugging.

Verification-aware languages, such as Dafny~\cite{dafny} and Verus~\cite{lattuada2023verus}, embed logical constructs, including preconditions (\emph{requires}), postconditions (\emph{ensures}), and invariants (\emph{invariant}), directly into code, allowing verifiers to automatically flag specification violations. While this is helpful in warning developers that the implementation does not match the expected behavior, it often does not indicate the source of the problem. In fact, a single failing verification condition may result from many possible fault locations, and the verifier feedback is often too coarse for debugging. This limitation is consistent with findings from prior work, which identified poor error messages and inadequate feedback as key challenges in the use of verification-aware languages, with users highlighting better debugging tools as an area for improvement \cite{oliveira2025challenges}. Currently, verifier feedback provides the violated condition and one representative counterexample path, but not a ranked explanation of which program lines are most likely responsible for the failure. 
As a result, developers are left to manually identify the root cause of the problem among many potential fault locations. This is especially problematic when multiple branches can satisfy the same failing proof obligation, because the first model found by the solver may point to a branch that is not the most useful to find the bug. 

This work investigates automated fault localization (FL) for verification-aware languages by comparing two %
paradigms:

\begin{itemize}
\item{ \textbf{State-based localization (SNAP)}: Based on an existing approach implemented in AutoFix ~\cite{AutoFix3} for Eiffel\cite{meyer1988eiffel}, which represents suspicious program states as ``snapshots'' (i.e., abstractions
of a concrete property of the program state at a concrete program location) and ranks them using a combination of static and dynamic evidence.}

\item \textbf{Counterexample localization}: We evaluate a family of counterexample-based localization techniques with increasing capabilities. These include raw baseline extraction (CNTB), single-trace structured ranking (CNTS), and, our primary contribution, a multi-trace aggregation via iterative path blocking (CNTM).
\end{itemize}

We additionally evaluate an LLM-based approach. However, because LLMs are neither deterministic nor verifier-native, this approach is included only as a baseline for comparison. %

A key challenge in our comparison is the architectural shift from dynamic to static verification. AutoFix was originally designed for Eiffel, where contracts are dynamically checked during program execution. In contrast, Dafny failures manifest as unsatisfied proof obligations during static analysis. 

We formalize the fault localization problem in verification-aware languages by treating any program that fails to verify as containing an implementation fault, i.e., a defect in executable code that violates a correct specification. While specification errors (e.g., weak postconditions) are common, this study isolates faults at the implementation level, allowing for a controlled evaluation with precise ground truth. Each fault localization strategy considered in this paper takes as input a faulty program and produces a ranking of the lines in that program according to \emph{suspiciousness}, i.e., from the most to least likely to contain the fault causing the specification violation. We evaluate the core techniques with a mutation-based framework built on DafnyBench\mbox{~\cite{loughridge2024dafnybench}} using MutDafny\mbox{~\cite{amaral2025mutdafny}}.

\smallskip
\noindent\noindent
{\bf Contributions.} 
The main contributions of this paper are:
\begin{itemize}
    \item A \textbf{unified framework} for systematic evaluation of fault localization approaches in verification-aware languages.
    \item \textbf{SNAP}: A state-based ranker adapted from the AutoFix approach that uses inferred predicates and snapshots for fault localization.
    \item \textbf{CNTM}: A multi-trace ranker that aggregates evidence across multiple diverse counterexamples generated through iterative path blocking.
    \item A \textbf{structural ranking heuristic} that prioritizes nesting depth and control-flow to isolate root causes from incidental execution steps.
    \item An \textbf{empirical study} characterizing trade-offs between symbolic techniques and LLMs, identifying structural awareness as a key driver for search efficiency.
\end{itemize}

\section{Snapshot State-Based Fault Localization}
\label{sec:SNAP}

We first present \textbf{SNAP}, a state-based approach that ranks program locations by identifying suspicious program states through the construction of a program state abstraction. SNAP is a Dafny adaptation of AutoFix's fault localization approach, originally proposed for Eiffel and the first designed specifically for verification-aware languages.%

Both languages are grounded in Design by Contract (DbC), but while Eiffel is a pure Object-Oriented (OO) programming language, Dafny has a more imperative character, where OO features are available but not the most commonly used ones. This distinction is significant because part of AutoFix's procedure for building a program abstraction uses the OO structure. Additionally, Eiffel supports DbC as a runtime mechanism, whereas Dafny statically verifies contracts before execution, so feedback regarding contract violations will differ in nature. The fundamental differences between the two languages mean that the same approach may yield different results. Studying these differences is therefore an important foundational step, establishing a baseline for further research on fault localization techniques for verification-aware languages.

SNAP is a dynamic analysis technique, relying on program executions to gather evidence about suspicious program states. This analysis is centered on the faulty method, defined as the method where Dafny's verifier reports a specification violation. AutoFix was designed for programs with single faults, so if the verifier flags multiple violations, SNAP handles only the first. Extending this approach to iteratively handle multiple violations is feasible but left as future work.

This approach investigates the possible causes of program faults by building an abstraction of the program state based on boolean predicates collected from multiple sources. Each of these boolean predicates creates a \emph{snapshot}, an abstraction of a concrete property of the program state at a concrete program location. Each snapshot is assigned a suspiciousness score derived from a combination of static and dynamic analysis features. The ranking of program lines produced by the strategy is projected from the snapshot ranking, considering the program location associated with each snapshot.%

While most fault localization techniques that compute suspiciousness scores do so at the program statement level, SNAP analyzes specific program states. This finer level of fault localization granularity provides more precise guidance for a future fix generation process. Its goal is to, given one of the top-ranked snapshots, attempt to avoid the state it abstracts.%

\subsection{Formal Model of Snapshots}

Let \(B\) be the collection of boolean predicates abstracting the program state, each predicate \(p \in B\) is evaluated at each location \(l\) of the program where it is visible, resulting in a snapshot, i.e., a triple $\langle l, p, v\rangle$, where \(v\) is a boolean value. 

Each snapshot abstracts a program execution in which \(p\) evaluates to \(v\) at location \(l\). A snapshot with a high suspiciousness score indicates there is a high probability of the fault being due to predicate \(p\) taking value \(v\) at location \(l\).

\subsection{Test Cases}

AutoFix relies on both contracts and tests automatically generated from the contracts as the repair oracle. There is considerable research on test generation from specifications~\cite{spec-based-testing1, spec-based-testing2, spec-based-testing3, spec-based-testing4}. Eiffel benefits from its own specification-based test generation tool, AutoTest~\cite{AutoTest}, which automatically selects inputs for the program constructs under test and determines their success or failure by observing whether they satisfy their contracts when executed with those inputs. However, no such approach exists in Dafny, which poses a challenge to the replication of this fault localization approach.

Dafny itself provides a beta version of a counterexample-based test generation command under development\footnote{\href{https://dafny.org/dafny/DafnyRef/DafnyRef\#sec-dafny-generate-tests}{https://dafny.org/dafny/DafnyRef/DafnyRef\#sec-dafny-generate-tests}, accessed Aug. 2026.}, resulting from the integration of the work developed first in Delfy~\cite{Delfy1, Delfy2} and subsequently in DTest~\cite{DTest}. However, this tool is insufficient for our needs because it does not support some of Dafny's key features, such as arrays, and it only generates one test case per program path. Our approach requires a broad and diverse test suite for detecting faulty execution patterns. Additionally, although it is guided by the specification, it generates tests based on the implementation's structure rather than independently from it, thus failing to distinguish between correct and incorrect implementation behavior.

To address this gap, we developed \textbf{DafnyCBT}, a contract-based test generator that \textit{derives both inputs and expected outputs from contracts}, independently of the implementation. This ensures that tests fail whenever the implementation deviates from its contract, and that specified behavior unreached by the code is still exercised.

Test generation has two stages. First, the contract is decomposed into Disjunctive Normal Form (DNF) using implication splitting that is safe under short-circuit evaluation, with each clause representing an input/output equivalence class. Z3 solves each satisfiable clause for concrete inputs and, where feasible, expected outputs. A uniqueness probe emits either a concrete pin \texttt{expect result==\textit{v}}, a disjunctive pin over a small enumerated set, or the postcondition literals as runtime \texttt{expect} assertions. Then, boundary value analysis targets collection sizes (e.g., \texttt{|a|}$\in\{0,1,{\geq}2\}$) and numeric extremes from relational literals, covering off-by-one faults. Two refinements improve fault detection: an \emph{anti-trivial bias} soft-asserted in each SMT query nudges Z3 away from trivial values that mask faults, and a literal \emph{relevance check} inspired by Modified Condition/Decision Coverage strengthens each clause query so every literal constrains admissible outputs.

A runtime check phase splits the suite into \texttt{Passing} and \texttt{Failing} methods, containing, respectively, the set of passing tests \(P\) and of failing tests \(F\), which are consumed by SNAP's fault localization procedure. 

Tool details and sources are available at \href{https://github.com/VeriFixer/DafnyCBT}{https://github.com/VeriFixer/DafnyCBT}.

\subsection{Snapshot Generation}

The set of boolean predicates from which snapshots are generated is identified by two different processes: (1) inference of invariants at each location of the faulty method, and (2) enumeration of boolean predicates relevant in the context of the faulty method.

\smallskip\subsubsection{Snapshot Generation via Invariant Inference}

An \emph{invariant} is a property that is always true at a given program location. For SNAP, the distinction between \emph{passing} and \emph{failing} invariants is fundamental in identifying faulty behavior. The set of \emph{passing invariants} \(\pi_l\) corresponds to properties that are true at program location \(l\) in every execution of the passing tests \(P\). Conversely, the set of \emph{failing invariants} \(\phi_l\) corresponds to properties that hold at location \(l\) for every execution of the failing tests \(F\). The boolean predicates in \(\Pi = \{p \: | \: p \in \phi_l \wedge p \notin \pi_l\}\), i.e., the properties that hold in failing tests but not in passing ones, are representative of potentially faulty behavior. For each invariant \(i \in \Pi\) inferred at location \textit{l}, a snapshot \(\langle l, i, true\rangle\) is generated.

We collect invariants using Daikon~\cite{Daikon}, a dynamic likely invariant inference tool. It is \emph{dynamic} because it operates by observing evidence of the program's behavior during execution, through the analysis of a trace file that records values of different program variables at different program locations. In turn, \emph{likely} means that the invariants are statistically likely to be true based on a group of observed executions. For some programming languages, Daikon provides frontends that automatically produce trace files from a program equipped with test cases. For the remaining languages, including Dafny, a custom frontend\footnote{\href{https://plse.cs.washington.edu/daikon/download/doc/developer.html\#New-front-ends}{https://plse.cs.washington.edu/daikon/download/doc/developer.html\#New-front-ends}, accessed Aug. 2026.} which converts execution data into a format that Daikon can process, must be provided.

Our frontend was implemented by instrumenting the input Dafny program with print statements. This way, running it and redirecting the output to a file produces a trace file ready to be fed into Daikon.
The first step is to print a header\footnote{More information about the expected header format is available at \href{https://plse.cs.washington.edu/daikon/download/doc/developer.html\#Declarations}{https://plse.cs.washington.edu/daikon/download/doc/developer.html\#Declarations}, accessed Aug. 2026.} consisting of program point declarations, i.e., the points of the program at which we want Daikon to perform invariant inference. In our case, these correspond to the faulty method's entry, exit, and every position between statements\footnote{By default, Daikon only supports invariant inference at methods' entries and exits, but the documentation mentions a workaround that can be achieved by inserting \textit{dummy} method calls, that take all visible variables as parameters, at the additional locations we wish to infer invariants at. The invariants produced by Daikon at the \textit{dummy} method's entry and exit are the invariants that hold at the call location.}. Each program point declaration contains a list of variable declarations for the variables visible at that program location, which we identify by analyzing the program prior to its instrumentation. The program traces are expressed after this initial block of declarations. Each trace record\footnote{For more information about the expected trace record format, visit \href{https://plse.cs.washington.edu/daikon/download/doc/developer.html\#Data-trace-records}{https://plse.cs.washington.edu/daikon/download/doc/developer.html\#Data-trace-records}, accessed Aug. 2026.} consists of the identification of one of the program points previously declared, and, for each of the variables declared in that point, its value in a given execution. We achieve this by printing the value of each visible variable at each location of the faulty method.

Since we want to generate passing and failing invariants separately, we perform two runs of this instrumentation process. Both runs insert the print statements required for trace generation. One additionally disables the failing tests by removing the invocation of \texttt{Failing}, while the other disables the passing tests by removing the invocation of \texttt{Passing}. As a result, in the first run, only passing executions are considered for invariant inference, while in the second, only failing executions are considered. %

\smallskip\subsubsection{\textcolor{black}{Snapshot Generation via Enumeration}}

SNAP complements the generation of snapshots via invariant inference with the enumeration of boolean predicates collected from different sources throughout the program, motivated by a study by Pytlik et al.~\cite{fault-localization-inv-study} that discusses the limitations of invariant inference for fault localization. Essentially, the invariants inferred by Daikon enable the discovery of more complex and semantically rich properties about the program, while the enumeration process provides completeness due to its exhaustive coverage of basic conditions.%

The enumeration process is, in principle, similar to invariant inference, except that it relies on expressions directly captured from the program and on simpler templates (adopted from AutoFix), rather than using those implemented by a tool, and then checks whether they hold during execution. The set of collected boolean predicates \(B\) is defined as follows:

\begin{itemize}
    \item \textbf{Boolean expressions}. For every boolean expression \(b\) appearing in the faulty method's contract or implementation, \(b \in B\).
    
    \item \textbf{Argumentless boolean queries (boolean-valued functions)}. For every argumentless boolean query \(q\) publicly available in the faulty method's scope, \(q() \in B\). 
    For Dafny, this corresponds to predicates and functions with boolean return type belonging to the same class as the faulty method
    (or scope, in case the program does not follow the OO paradigm, as often happens in Dafny).
    
    \item \textbf{Constructed argumentless boolean queries.} For every object \(o\) visible in the faulty method's contract or implementation, and every argumentless boolean query \(q\) publicly available in \(o\)'s class, \(o.q() \in B\).

    \item \textbf{Boolean expressions constructed from integer expressions.} For every two integer expressions \(e\) and \(e'\) appearing in the faulty method's contract or implementation, or resulting from calling each available argumentless integer query on each object appearing in the faulty method's contract or body, \(\{e = e', e < e', e \leq e'\} \subset B\) (and, by symmetry, \(\{e' < e, e' \leq e\} \subset B\)), as well as, \(\{e = 0, e < 0, e \leq 0\} \subset B\) (and, by symmetry, \(\{e' = 0, e' < 0, e' \leq 0\} \subset B\)).%
    
    \item \textbf{Null checks.} For every nullable reference type expression \(e\) appearing in the faulty method's contract or implementation, or resulting from calling each available argumentless nullable reference type query on each object appearing in the faulty method's contract or body, \(e = null \: \in B\).
    
    \item \textbf{Logical implications.} For each logical implication \(a \implies b\) appearing in the faulty method's contract or in the contract of the included queries, \( \{a \implies b, \neg a \implies b, a \implies \neg b, b \implies a\} \subset B\). 

\end{itemize}

For each program location \(l\), and each boolean predicate \(p \: \in B\) capable of being evaluated at \(l\) (i.e., every variable appearing in sub-expressions of \(p\) is visible at \(l\)), a snapshot \(\langle l, p, v\rangle\) is generated, where the value of \(v\) is either \(true\) or \(false\), as dynamically observed during test suite execution. The triple \(\langle l, p, true\rangle\) will be included in the set of snapshots if predicate \(p\) is \(true\) at location \(l\) in at least one test execution. The same applies to \(\langle l, p, false\rangle\).

\subsection{Suspiciousness Score Computation}\label{sec:suspiciousness}

A suspiciousness score is a metric commonly used in fault localization techniques to estimate how likely a program component is to be the cause of a fault. Several formulas for the computation of this metric have been proposed in the literature~\cite{tarantula1, sbfl1, DStar}. SNAP's formula combines three factors. Two static metrics are taken into consideration for calculating the suspiciousness score of a snapshot. The first is \emph{control dependence (cd)}, i.e., the distance between the snapshot's location and the location where the contract violation occurs when taking the shortest direct path. The second is \emph{expression dependence (ed)}, i.e., the similarity between the snapshot's predicate and the violated contract in terms of how many sub-expressions they share. The dynamic analysis metric \emph{(dyn)} consists of a weighted sum measuring how often the snapshot appears in failing test runs as opposed to passing ones. Its value is higher the more it appears in failing tests and the less it appears in passing tests, following spectrum-based fault localization approaches~\cite{sbfl1, sbfl2}. 


Each of the three metrics, as well as the overall suspiciousness score for a snapshot \(\langle l, p, v\rangle\), is, respectively, computed as follows:

\noindent\noindent
\begin{equation*}
cd(l, f) = 1 - \frac{cdist(l, f)}{max\{cdist(\lambda, f) \: | \:\lambda \in m \wedge \lambda \rightsquigarrow f\}}
\end{equation*}
where \(m\) is the faulty method, \(f\) is the location where the contract violation occurs, \(l_1 \rightsquigarrow l_2\) represents the existence of a direct path from program location \(l_1\) to program location \(l_2\), and \(cdist(l_1, l_2)\) represents the length of the shortest path from \(l_1\) to \(l_2\).

\noindent\noindent
\begin{equation*}
ed(p, c) = \frac{eprox(p, c)}{max\{eprox(\lambda, c) \: | \: \lambda \in B \}}
\end{equation*}

where \(c\) is the violated contract, and \(eprox(e_1, e_2)\) represents the number of shared sub-expressions between expression \(e_1\) and expression \(e_2\).
\noindent\noindent
\begin{equation*}
dyn(\langle l, p, v\rangle) = \gamma + \dfrac{\alpha}{1 - \alpha} (1 - \beta + \beta\alpha^{\#p(\langle l, p, v\rangle)}) - \alpha^{\#f(\langle l, p, v\rangle)}
\end{equation*}
where \(\#p(\langle l, p, v\rangle)\) is the number of passing tests where the snapshot appears, \(\#f(\langle l, p, v\rangle)\) is the number of failing tests where the snapshot appears, and \(\alpha\), \(\beta\), and \(\gamma\) are non-negative real numbers. AutoFix empirically determined \(\alpha = 1/3\), \(\beta = 2/3\), and \(\gamma = 1\), which we adopt in our implementation.
\noindent\noindent
\begin{equation*}
susp(\langle l, p, v \rangle) = \dfrac{3}{cd(l, f)^{-1} + ed(p, c)^{-1} + dyn(\langle l, p, v\rangle)^{-1}} 
\end{equation*}
This suspiciousness score formula takes the dynamic score as the primary source of evidence, and control and expression dependence as secondary criteria to discriminate between snapshots with similar dynamic score (dynamic score has a minimum of \(1\) while the others vary between \(0\) and \(1\)).

Calculating the dynamic score involves knowing how many times a snapshot appears in executions of passing and failing tests. We collect this information by instrumenting the faulty program. For each \(\langle l, p\rangle\) snapshot pair, we insert one print statement evaluating \(p\) at location \(l\) of the faulty method. The output of the instrumented program allows us to count how many times each snapshot appears across test executions.

\section{Counterexample-Based Fault Localization}
\label{sec:counterexample-techniques}

Counterexample-based techniques use the diagnostic information produced by the SMT solver when a verification condition fails. These methods transform the abstract execution traces (counterexamples) into a ranked list of suspicious program lines. We categorize these techniques into three levels of increasing complexity: \textit{baseline} (CNTB), \textit{structured} (CNTS), and \textit{multi-trace} (CNTM). CNTM is the primary technique proposed in this work, while CNTB and CNTS are simplified variants that isolate the contribution of its components.

\paragraph{Counterexample Baseline (CNTB)}

\textbf{CNTB} extracts counterexample data directly from the diagnostic messages produced by the Dafny CLI when run with the \texttt{extract-counterexample} flag. This method produces exactly one counterexample for each failing verification condition (e.g., an assertion or postcondition), providing a single witness even if multiple failing paths exist for that same condition. CNTB performs no advanced heuristic weighting, it simply identifies every line involved in the reported execution path and maps it to the corresponding source code, resulting in a raw set of suspicious lines.

\paragraph{Counterexample Single-Trace Structured Ranker (CNTS)}

\textbf{CNTS} significantly refines the raw output of the baseline by introducing code-awareness in a ranking heuristic that prioritizes program lines based on their structural properties. CNTS's two primary functional improvements over CNTB consist of its enhanced structural extraction and formal ranking system. Because the tool is code-aware, it extends the diagnostic set beyond the literal lines found in the trace. It identifies and flags as suspicious the condition nodes of \texttt{if}, \texttt{while}, and \texttt{for} blocks whenever a line identified in the failing trace exists within their scope. Additionally, CNTS provides discriminatory power that the baseline lacks by ranking these lines using structural heuristics based on line frequency, nesting depth, and node type. %

\paragraph{Counterexample Multi-Trace Aggregator (CNTM)}

\textbf{CNTM} addresses the primary limitation of the single-trace approach in that a single counterexample is merely one of many possible ways a program can fail. Relying on a single trace per verification condition failure can lead to an incomplete exploration of the fault space. 
CNTM improves upon CNTS by forcing SMT solver diversity through an iterative path blocking mechanism. Upon finding a counterexample, the system identifies the associated execution path and introduces a temporary constraint to block that specific path in the solver's next search. This process is repeated to collect multiple, independent failing paths. CNTM then aggregates evidence across all collected traces, assigning higher suspiciousness to lines that appear consistently across diverse failure paths.

\subsection{Formal Model of Counterexamples}

A verification failure produces a counterexample \(C\), formalized as a tuple $\langle \sigma, \pi \rangle$, where $\sigma$ represents the \emph{valuation of variables} and $\pi = \{ \ell_1, \ell_2, \dots, \ell_n \}$ the \emph{trace path} sequence, defined as the ordered program lines executed in the counterexample.

\smallskip\subsubsection{Path Independence}
By default, Dafny only provides one counterexample trace per failed verification condition. To improve coverage, CNTM seeks a set of \emph{independent counterexamples}. Two counterexamples \(C_i = \langle \sigma_i, \pi_i \rangle\) and \(C_j = \langle \sigma_j, \pi_j \rangle\) are \emph{path-independent} if and only if their execution paths differ: $\pi_i \neq \pi_j$. This distinction ensures that we prioritize discovering new buggy branches over exploring different data valuations of the same logical error. The aggregate path set $\Pi = \{\pi_i, \pi_j, ..., \pi_n\}$ includes all the collected independent trace paths.

\subsection{Iterative Path Extraction Algorithm (CNTM)}

To address the limitation of the Dafny verifier providing a single counterexample path per failing condition, we implement an iterative instrumentation strategy to exhaustively explore alternative independent failing execution traces. The algorithm prunes previously discovered paths, forcing the SMT solver to find different ways the same verification condition might fail. For a failing verification condition $v$ (e.g., an assertion or postcondition), the process is as follows:

\begin{enumerate}
    \item \textbf{Initial Extraction.} The verifier runs to obtain an initial counterexample $C$ for condition $v$, and its associated execution path $\pi$ from the Dafny CLI diagnostic messages.
    \item \textbf{Path Pruning.} Identification of the \emph{deepest branch point} within $\pi$, representing the final logical decision (e.g., an \texttt{if} branch or loop entry) that led to the failure.
    \item \textbf{Instrumentation.} Temporary instrumentation of the source code through the injection of an \texttt{assume false} statement at this specific branch point, effectively pruning the discovered path from the search space.
    \item \textbf{Iterative Discovery.} The verifier is re-run on the instrumented version of the code. If a new counterexample $C'$ for $v$ is identified by the solver, the new path $\pi'$ is extracted and the process is repeated from Step 2.
    \item \textbf{Termination.} The loop terminates when condition $v$ either verifies or no further counterexamples are generated, indicating all unique failing paths have been captured.
\end{enumerate}

\subsection{Multi-Criteria Ranking Across Counterexamples (CNTS, CNTM)}
\label{subsec:ranking}

Once the aggregate path set $\Pi$ is collected, we rank each line $l \in \bigcup \Pi$, where $\bigcup \Pi$ denotes the set of all unique program lines appearing across the paths in $\Pi$, using a hierarchical scoring system to prioritize root causes over incidental execution steps. This ensures that the most suspicious lines are presented to the developer first. The ranking is determined by the lexicographical comparison of the following tuple:
\begin{equation*}
\text{rank}(l) = \langle  -\text{freq}(l), -\text{max\_dep}(l), -\text{is\_cntr}(l), \text{ord}(l) \rangle
\end{equation*}

This order is defined by four metrics:
\begin{itemize}
    \item \textbf{Frequency} ($\text{freq}(l)$): The number of independent paths in $\Pi$ that contain line $l$ serves as the primary filter. Lines recurring across multiple independent paths represent a ``consensus'' of suspicion.
    \item \textbf{Maximum Depth} ($\text{max\_dep}(l)$): The maximum nesting level of the statement at line $l$. Targets structural complexity, based on the heuristic that logical errors frequently hide within deeply nested blocks.
    \item \textbf{Control Statement} ($\text{is\_cntr}(l)$): A boolean indicator of whether line $l$ consists of a branch guard (e.g., \texttt{if/while/for}). These are prioritized to highlight the \emph{decisions}, serving as a tie-break if other lines are at the same nesting level as these control statements.
    \item \textbf{Trace Order} ($\text{ord}(l)$): The absolute position of line $l$ in the file, used only to break ties.
\end{itemize}

\subsection{Multi-Trace Aggregator Execution Example (CNTM)}

To illustrate the iterative path extraction process, consider the \texttt{demo} method in Listing~\ref{lst:demo-bugs}, which contains two distinct logical errors, in Lines \ref{line:bug1} and \ref{line:bug2}, that violate the postcondition \(y = x\) for positive inputs.

\begin{lstlisting}[style=dafnystyle, caption={Example of path extraction via instrumentation, used by CNTM.}, captionpos=b, aboveskip=10pt, label={lst:demo-bugs}, escapeinside={(*}{*)}]
method demo(x: int) returns (y: int)
  ensures x > 0 ==> y == x { 
  if x > 1 {           // Path C2: (x=2) 
    y := x + 1;        // Bug: y should be x(*\label{line:bug1}*)
    return y;
  } if x > 0 {         // Path C1: (x=1)
    // assume false(*\label{line:instrumentation}*)
    y := x + 2;        // Bug: y should be x(*\label{line:bug2}*)
    return y;
  }
  return x;
}
\end{lstlisting}

For this example, the algorithm proceeds as follows: 

\smallskip\noindent\noindent
\emph{\textbf{1) Initial Counterexample Extraction ($C_1$).}} Upon the first verification attempt, the SMT solver identifies a counterexample where $x=1$. The execution path $\pi_1$ skips the first \texttt{if} block (since $1 \ngtr 1$) and enters the second \texttt{if} block ($1 > 0$). The solver returns the trace $\pi_1 = \{\text{line 6, line 8, line 9}\}$, where Line \ref{line:bug2} contains the buggy assignment \texttt{y := x + 2}.

\smallskip\noindent\noindent
\emph{\textbf{2) Path Pruning via Instrumentation.}} To discover other potential faults, the algorithm instruments the second \texttt{if} branch. As shown in Line \ref{line:instrumentation}, an \texttt{assume false} statement is dynamically injected at the start of the branch explored by $C_1$. This statement instructs the solver to ignore any execution path that enters this specific block during the next iteration.

\smallskip\noindent\noindent
\emph{\textbf{3) Second Counterexample Extraction ($C_2$).}} With the path $\pi_1$ effectively blocked, the verifier is forced to explore alternative branches to satisfy the negation of the postcondition, identifying $x=2$. This execution path $\pi_2$ explores the first \texttt{if} block ($2 > 1$), yielding $\pi_2 = \{\text{line 3, line 4, line 5}\}$, where Line \ref{line:bug1} contains the second buggy assignment \texttt{y := x + 1}. Since, after blocking these two branches, no further failing paths exist, the iteration terminates.

\smallskip\noindent\noindent
\emph{\textbf{Resulting Ranking Input.}} The algorithm concludes with the aggregate set $\Pi = \{ \pi_1, \pi_2 \}$. By forcing the solver to move past the first discovered error, we ensure that both \texttt{if} blocks are identified as candidates for fault localization. Note that if a line were shared between these two independent paths, its Frequency count would increase, correctly marking it as a highly suspicious common point of failure. %

\label{subsec:rankingdeleted}

 \section{LLM-Based Fault Localization}
\label{sec:llm-techniques}

As a complement to the verifier-native techniques presented above, we present \textbf{LLM}, a fault localization baseline that relies on a large language model to rank suspicious program lines. Given a verification failure, the model receives the numbered source code and is asked to identify the most likely line(s) responsible for the failure. In this baseline, the model receives no additional information.

Our goal is not to optimize LLM performance, but rather to compare verifier-native fault localization techniques against a representative LLM-based approach. Fully optimizing the LLM would introduce additional experimental variables, such as prompt engineering, model selection, retrieval mechanisms, and few-shot examples, requiring dedicated ablation studies beyond the scope of this work. Still, to strengthen the comparison, our study includes two additional variants. \textbf{LLM\_ER} augments the original setup with the verifier error message, and \textbf{LLM\_ER\_CN} constitutes a hybrid configuration combining verifier error messages with CNTM's ranking, aiming to capture potential synergies between the two. We used the following prompt:

\vskip 0.5em
\begin{small}
\begin{Verbatim}[fontsize=\footnotesize]
You are a fault localization model. Rank suspicious
lines from most to least likely to be faulty. 
Output contract (strict):
1) The list is a ranking: the first element must 
be the line most likely to contain the fault.
2) You do not have to return a full list with all 
the lines, prefer shorter lists.
3) Return exactly one JSON array of unique 1-based 
line numbers.
4) Do not return any explanation, markdown, code 
fences, labels, or extra text.
5) Use integers in the inclusive range[1, t_lines].
6) If no suspicious lines are found, return [].
7) Your entire response must match this pattern:
^\\[(?:\\s*\\d+\\s*(?:,\\s*\\d+\\s*)*)?\\]$
Example valid response: [30, 15, 29]
{verifier error if LLM_ER variant}
The file has {t_lines} lines.
BEGIN_FILE
{numbered_source_code}
END_FILE
\end{Verbatim}
\end{small}
\vskip 0.5em

We used \texttt{qwen3-coder-next} model via OpenRouter (tested in April 2026). Generation followed OpenRouter's defaults, with temperature 1.0, top\_p 0.95, top\_k 40, a maximum output length of 66,000 tokens, and a context window of 262,144 tokens. We selected it to balance reproducibility, performance, and cost. Unlike proprietary models, which change over time and hinder exact replication, open-weight models provide fixed, versioned checkpoints that enable reproducibility. At the time of this study, the model was the latest Qwen code model and among the strongest open-weight coding LLMs available, achieving competitive results on coding benchmarks (e.g., 70\% on SWE-bench \cite{cao2026qwen3}).

\section{Evaluation}\label{sec:evaluation}

\subsection{Research questions}

Our empirical evaluation aims to answer the following research questions: \\
\textbf{RQ1 (State vs. Path):} How do state-based FL techniques compare to counterexample-based FL techniques? \\
\textbf{RQ2 (Ranking Ablation Study):} How do the individual ranking components (frequency, depth, and structure) compare in terms of their contribution to FL effectiveness? \\
\textbf{RQ3 (Path Diversity):} How does multi-trace aggregation compare to single-trace analysis in terms of FL effectiveness? \\
\textbf{RQ4 (LLM Baseline Comparison):} How do verifier-native FL techniques compare to a general-purpose LLM baseline?

\subsection{Experimental Procedure}

We evaluate SNAP, the counterexample techniques (CNTB, CNTS, and CNTM), and the LLM techniques (LLM, LLM\_ER, and LLM\_ER\_CN) within a unified experimental framework on Dafny mutants derived from DafnyBench. As a reference point, we include a \textbf{RAND} baseline, which returns all lines of the failing method randomized, serving as a sanity check for calibration rather than as a competing technique.

We generated our evaluation benchmark using MutDafny \cite{amaral2025mutdafny} and all of its mutation operators to produce mutants from the 785 programs of DafnyBench \cite{loughridge2024dafnybench}. The set of generated mutants includes 68,957 that fail verification, i.e., contain faults detected by Dafny's verifier. However, MutDafny stores mutants in their post-resolution representation and, when converting them back to source code, this process can produce invalid Dafny programs. We, therefore, filtered for valid, compilable code, reducing the set to 57,925 mutants.

\begin{table}[ht]
\caption{Mutation operators used in the artificial fault dataset. \#M is the number of mutants generated with each operator.}
\centering
\begin{tabular}{p{0.03\columnwidth}p{0.7\columnwidth}p{0.03\columnwidth}}
\hline
\textbf{Op.} & \textbf{Description} & \textbf{\#M} \\
\hline
AOR & Replacement of an arithmetic operator with another & 33 \\
ROR & Replacement of a relational operator with another & 51 \\
COR & Replacement of a conditional operator with another & 4 \\
BBR & Replacement of a relational or conditional expression with \texttt{true} or \texttt{false} & 20 \\
AOI & Insertion of a unary minus in an arithmetic expression & 76 \\
COI & Insertion of a \texttt{not} operator in a conditional expression & 21 \\
AOD & Deletion of a unary minus in an arithmetic expression & 1 \\
COD & Deletion of a \texttt{not} operator in a conditional expression & 1 \\
LVR & Replacement of a numerical literal with its increment, decrement, or zero, and of a string literal with an empty one, a default one, or a mutation of the original & 50 \\
EVR & Replacement of an expression with a default literal & 62 \\
VER & Replacement of same-type variables & 92 \\
LBI & Insertion of a \texttt{break} statement at the beggining of the body of a loop & 5 \\
MRR & Replacement of a call with a default return-type literal & 6 \\
MAP & Replacement of a call with one of its arguments with the same type as the return value & 1 \\
MVR & Replacement of a call with a same-type variable & 8 \\
SAR & Swap of a call argument with a same-type one used in the same call & 2 \\
CIR & Replacement of a non-empty collection initializer with an empty one, and of an empty initializer with a default non-empty one & 6 \\
CBE & Extraction of one of the blocks of an if or if-then-else statement and deletion of the remaining ones & 7 \\
DCR & Replacement of a datatype constructor with another of the same datatype and with the same signature & 1 \\
FAR & Replacement of a class's field access with another same-class field & 1 \\
SDL & Deletion of a statement or of an entire code block & 28 \\
ODL & Deletion of all occurrences of a binary operator (and of one of its arguments to preserve program validity) & 12 \\
SWS & Swap of a statement with a neighboring one & 12 \\
\hline
\end{tabular}
\label{tab:mutation-operators-dataset-distribution}
\end{table}

Given the high computational cost of iterative SMT solver calls and invariant inference, we randomly sampled 500 mutants for processing. This sample size is statistically robust, as studies suggest that 300 to 1,000 mutants are sufficient to yield stable fault localization results and approximate mutation scores within narrow error bounds \cite{gopinath2015hard, steimann2013threats}.

Our dataset of 500 programs comprises a total of 44,242 lines of code, averaging 88 lines per program file. Additionally, on average, each program contains five non-ghost components, which in Dafny includes methods, functions, and other similar constructs. \Cref{tab:mutation-operators-dataset-distribution} summarizes the 23 mutation operators, out of 40 supported by MutDafny, that led to the generation of the 500 mutants included in our dataset, showcasing the variety of types of faults it tries to mimic. These include 16 mutation operators previously proposed in mutation tools for other programming languages, as well as seven newly proposed by MutDafny. These seven operators were selected based on an analysis of bugfix evidence from Dafny projects available on GitHub, making the mutants they generate representative of real mistakes Dafny developers make.

For each mutant, the ground truth, i.e., the fault location, was established by performing a diff against the original program, to identify the mutated location.

Our experimental procedure involves applying each fault localization technique to each of the 500 faulty programs in our dataset. For each strategy and mutant, we produce a ranked list of lines \(L = [l_1, \dots, l_t]\), ordered by suspiciousness, and evaluate it against the ground-truth faulty line \(l_{gt}\).

\subsection{Experimental Metrics}

We evaluate fault localization effectiveness using the EXAM score, which captures the effort required to locate the faulty line, Top-$k$ success rates, and two coverage indicators: \emph{Found} rate, measuring the proportion of cases where the ground truth appears in the model's ranking, and \emph{Empty} rate, representing the proportion of cases where the model returns no lines.

For the EXAM score, we adopt a normalized, 0-based formulation. Let \(N\) denote the number of candidate lines in the search scope, and let \(r\) be the number of non-fault lines appearing in the ranked list \(L\) inspected before matching the ground-truth faulty line \(l_{gt}\). It is defined as $EXAM = r/(N - 1)$, where a value of 0 indicates immediate localization and 1 indicates that the fault is found last in the ranking.

Rankings are computed over absolute line numbers within the active scope (file or method). However, strategies are not required to return rankings containing every single line. When the faulty line is absent from an output of size \(t\), we estimate the remaining effort by assuming that the developer continues inspecting the remaining \(N - t\) lines in random order. Under this assumption, the fault is expected to appear at the midpoint of the unranked portion, and the EXAM score is computed accordingly, as proposed by Pearson et al. \cite{pearson2017evaluating}.

We assess statistical significance on EXAM scores using paired, two-sided Wilcoxon signed-rank tests. We report \emph{p-values} together with matched rank-biserial correlations as effect sizes. For Top-1 localization, we use McNemar's exact test and report paired odds ratios.

\subsection{Dual-Scope Evaluation}

We perform our empirical evaluation under two scopes, capturing two different debugging scenarios:
\begin{itemize}
  \item \textbf{File scope}: Considers all executable lines in the program file, reflecting the standard fault localization setting where there are no assumptions about the fault location \cite{pearson2017evaluating}.
  \item \textbf{Method scope}: Considers executable lines within the faulty method, reflecting a workflow where the developer has already narrowed the search space using the verifier's error message. This evaluates localization performance within the identified faulty method.
\end{itemize}

\subsection{Comparison with State-of-the-Art Methods} We focus on verifier-native methods for verification-aware programming languages, where specifications can provide a more reliable correctness oracle than test suites. As a result, our evaluation does not include a comparison to classical fault localization methods (e.g., spectrum-based (SBFL) and mutation-based (MBFL)) that rely solely on test suites as the correctness oracle. Porting test-based methods to Dafny would require a higher-quality test generator capable of producing a diverse suite with reliable pass/fail labels, a capability Dafny and even our DafnyCBT prototype currently lack. Because the resulting effectiveness would be dominated by test-generation quality rather than by the localization technique itself, a fair evaluation would need to distinguish these two factors. We therefore consider a systematic comparison of test versus specification oracles for fault localization to be a distinct research contribution, outside the scope of this paper.

\section{Results}
\label{sec:results}

\begin{figure}[t]
  \centering
  \includegraphics[width=0.95\columnwidth]{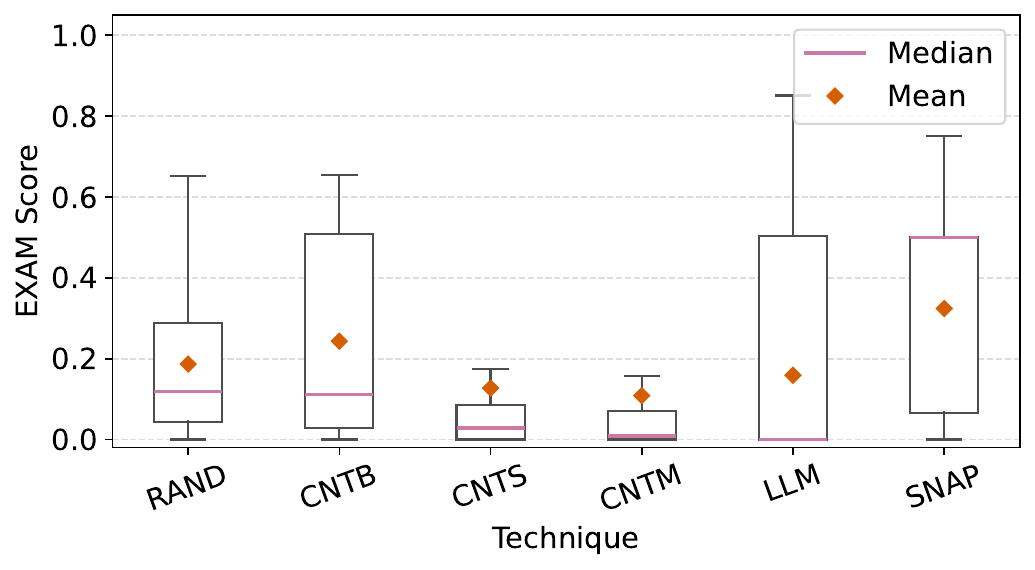}
  \caption{Distribution of file-scoped EXAM scores. The boxplots illustrate the median, quartiles, and outlier variability across techniques. Boxplots use Tukey whiskers (1.5×IQR); outliers are not shown.
  } 
  \label{fig:file-boxplot}
\end{figure}

Tables~\ref{tab:rq1_all}-\ref{tab:rq4_all} contain the scores for the different metrics for all the fault localization models. Figure\mbox{\ref{fig:file-boxplot}} illustrates the distribution of results.

For EXAM, CNTM significantly outperforms all non-LLM baselines ($p < 0.05$), while differences with LLM are not significant ($p = 0.101$) and have a negligible effect size. For Top-1, LLM significantly outperforms CNTM ($p < 0.05$), with CNTM outperforming all other baselines. Thus, the answers to the research questions focus more on CNTM than on the other counterexample-based variants. Full statistical results are provided in the Appendix in Tables \ref{tab:pairwise_wilcoxon_file} and \ref{tab:pairwise_mcnemar_top1_file}.

\subsection{\textbf{RQ1 (State vs. Path):} How do state-based FL techniques compare to counterexample-based FL techniques?}

Table~\ref{tab:rq1_all} shows that our counterexample-based approach outperforms the state-based approach we adapted from the literature. At file scope, SNAP achieves an EXAM score of 0.324, compared to 0.109 for CNTM, with a similar gap at method scope. One key factor is SNAP's high rate of empty predictions (39.20\%), which are penalized by the EXAM metric (assigned a default score of 0.5 when no ranking is produced). 
This high \emph{Empty} rate is mostly due to limitations of the testing tool, which, as an early prototype, produces valid test files for only 387 out of the 500 programs in our dataset, lacking support for programs involving more complex language features, such as multidimensional arrays, function arguments, and nested types (e.g., \texttt{map<int, (int, int)>}). Additionally, in 61 programs, though valid tests are generated, they either do not exercise the faulty method or fail to split the tests into passing and failing, a prerequisite for SNAP. From the remaining, 11 timed out after six hours, and another 11 failed after exhausting the available memory, since the state-based technique is also significantly more computationally expensive than the others. For the 297 programs for which all techniques produce non-empty outputs, SNAP attains an EXAM score of 0.212 at file scope and 0.359 at method scope, which is still substantially worse than CNTM. 

In the file setting, SNAP performs worse than RAND, which merely ranks lines uniformly at random among failing methods. Nevertheless, SNAP performs better than RAND when evaluated at the method scope, highlighting RAND's distinct behavior across scopes. RAND restricts its search to the lines of the failing method, and, because the method represents only a fraction of the file, the expected file-scope EXAM for RAND is approximately the method-to-file size ratio (around 0.19 in our dataset). Consequently, any technique that merely identifies the correct method already approaches this baseline. However, at method scope, RAND yields a score near 0.5 (0.483), representative of random line selection. The fact that SNAP outperforms RAND at this scope (0.359) indicates that, when the correct method has already been isolated, SNAP's snapshot-based heuristics do carry a meaningful signal for distinguishing faulty statements from correct ones. Despite this, the results still indicate that SNAP's limitations go beyond incomplete support for certain classes of programs. Even when it produces outputs, its ranking quality remains subpar.

This discrepancy is also reflected in effectiveness. In addition to SNAP's high \emph{Empty} rate (39.20\%), it only locates 47.00\% of faults, while CNTM achieves a \emph{Found} rate of 84\% with only 1.4\% of empty outputs. 
A likely contributing factor is SNAP's reliance on test executions, its effectiveness being tied to how well the generated tests exercise relevant program behavior. When paths containing faults are insufficiently covered, SNAP may fail to produce meaningful rankings or any output at all. In contrast, counterexample-based techniques operate directly on verifier-provided counterexamples that encode failure-inducing executions, without requiring test generation. As a result, they are less sensitive to coverage limitations.
Overall, SNAP's weaker performance appears to result from both limited coverage in test-based execution and lower-quality rankings, even when outputs are available.

\begin{table}[t]
    \caption{Fault localization on \textbf{All} (500 programs) and \textbf{Comp.} (the 297 with non-empty outputs from all methods). Metrics: EX$_F$/EX$_M$ (EXAM score, file/method scope); Top-$k$ (\% of faults ranked within the top k predictions); F (\% of cases that successfully locate the fault); E (\% that return an empty output).}
    
    \centering
    \small
    \begin{tabular}{llcccc}
        \toprule
        \textbf{Data} & \textbf{Strat.} & \textbf{EX$_F$} & \textbf{EX$_M$} & \textbf{Top-1/3/5} & \textbf{F / E} \\
        \midrule
        \multirow{3}{*}{All}
        & CNTM & 0.109 & 0.155 & 43 / 71 / 78 & 84.00 / 1.40 \\

        & SNAP & 0.324 & 0.413 & \mbox{\phantom{0}8} / 20 / 32 & 47.00 / 39.20 \\
        & RAND & 0.187 & 0.483 & \phantom{0}6 / 22 / 35 & 95.40 / 1.00 \\
        \midrule
        \multirow{3}{*}{Comp.}

        & CNTM & 0.092 & 0.138 & 46 / 77 / 83 & 87.54 / 0.00 \\  

        & SNAP & 0.212 & 0.359 & 11 / 33 / 53 & 77.44 / 0.00 \\

        & RAND & 0.186 & 0.479 & \phantom{0}7 / 26 / 39 & 98.99 / 0.00 \\
        \bottomrule
    \end{tabular}
    \label{tab:rq1_all}
\end{table}

\begin{tcolorbox}[myboxstyle]
Our novel counterexample-based technique largely outperforms the state-based technique, achieving higher overall effectiveness across a larger number of programs.
\end{tcolorbox}

\subsection{\textbf{RQ2 (Ranking Ablation Study):} How do the individual ranking components (frequency, depth, and structure) compare in terms of their contribution to FL effectiveness?}

To isolate the contribution of each ranking component, we perform an ablation study on the CNTM approach and four variants. CNTM\_RAW uses only raw CLI counterexample lines without augmenting them with control-flow data, and the three remaining versions respectively ignore \emph{frequency} (CNTM\_NF), \emph{depth} (CNTM\_ND), and \emph{control-statement} (CNTM\_NC) priority in the ranking formula.

The results, summarized in Table \ref{tab:rq2_all}, demonstrate that structural awareness is a significant driver of localization precision. The most significant performance drop occurs in the CNTM\_RAW variant. By discarding the augmented logical context, the EXAM score degrades dramatically (from 0.109 to 0.190 at file scope), meaning a developer would have to search almost twice as much code to find the fault. Furthermore, the Top-1 accuracy collapses from 43\% to just 18\%.

\begin{table}[t]  
    \caption{Fault localization performance across strategies (500 programs). Metrics as defined in Table~\ref{tab:rq1_all}.}
    \centering
    \small
    \begin{tabular}{lcccc}
        \toprule
        \textbf{Strat.} & \textbf{EX$_F$} & \textbf{EX$_M$} & \textbf{Top-1/3/5} & \textbf{F / E} \\
        \midrule
        CNTM & 0.109 & 0.155 & 43 / 71 / 78 & 84.00 / 1.40 \\
        CNTM\_RAW & 0.190 & 0.299 & 18 / 39 / 58 & 74.40 / 1.40 \\
        CNTM\_NF & 0.109 & 0.160 & 36 / 71 / 79 & 84.00 / 1.40 \\
        CNTM\_ND & 0.139 & 0.244 & 22 / 49 / 64 & 84.00 / 1.40 \\
        CNTM\_NC & 0.113 & 0.166 & 37 / 68 / 77 & 84.00 / 1.40 \\
        \bottomrule
    \end{tabular}
    \label{tab:rq2_all}
\end{table}

Among the individual ranking heuristics, \emph{depth} proved to be the most critical. Removing it (CNTM\_ND) caused the EXAM score to worsen significantly (rising to 0.139 at file scope and 0.244 at method scope) and slashed Top-1 performance by half (22\%). This confirms that the hierarchical position of a statement is a high-signal indicator of a fault's root cause.

The suppression of both the \emph{frequency} (CNTM\_NF) and \emph{control-flow} (CTNM\_NC) heuristics did not cause any statistically significant EXAM score changes regarding the baseline. However, in Top-1 accuracy, there is a noticeable dip from 43\% to, respectively, 36\% and 37\%. This suggests that, while these factors help pinpoint the exact bug as the first result, structural features like depth are more vital for reducing the overall search space.

\begin{tcolorbox}[myboxstyle]
 Structured ranking is a major contributor to localization quality. Neglecting structural context doubles the developer's search effort (EXAM) and reduces Top-1 accuracy.
\end{tcolorbox}

\subsection{ \textbf{RQ3 (Path Diversity):} How does multi-trace aggregation compare to single-trace analysis in terms of FL effectiveness?}

\begin{table}[t]
    \caption{Fault localization performance across strategies (500 programns). Metrics as defines in ~\ref{tab:rq1_all}}
    \centering
    \small
    \begin{tabular}{lcccc}
        \toprule
        \textbf{Strat.} & \textbf{EX$_F$} & \textbf{EX$_M$} & \textbf{Top-1/3/5} & \textbf{F / E} \\
        \midrule
        CNTB & 0.243 & 0.322 & 12 / 31 / 50 & 61.40 / 5.20 \\
        CNTS & 0.127 & 0.181 & 27 / 64 / 76 & 80.80 / 2.60 \\
        CNTM & 0.109 & 0.155 & 43 / 71 / 78 & 84.00 / 1.40 \\
        \bottomrule
    \end{tabular}
    \label{tab:cnt_subset}
\end{table}

Compared to CNTB and CNTS, CNTM consistently improves across all metrics, as depicted in Table \ref{tab:cnt_subset}. CNTB's performance is uniformly lower and, thus, not further discussed. For CNTS, at file scope, EXAM decreases from 0.127 to 0.109 with significance but moderate effect, and, at method scope, from 0.181 to 0.155. Figure~\ref{fig:file-boxplot} shows that CNTM outperforms CNTS, exhibiting a lower median and an overall distribution shifted closer to zero, indicating it tends to discover faults earlier across the evaluated budget range.

However, the gains are better reflected in Top-$k$ performance, with Top-1 improving substantially from 27\% (CNTS) to 43\% (CNTM), and with corresponding improvements in Top-3 and Top-5 accuracy. Coverage also improves, with the \emph{Found} rate increasing from 80.80\% to 84.00\%, and the \emph{Empty} rate decreasing from 2.60\% to 1.40\%.

\begin{tcolorbox}[myboxstyle]
Using diverse counterexample paths, multi-trace aggregation significantly improves ranking quality and robustness.
\end{tcolorbox}

\subsection{\textbf{RQ4 (LLM Baseline Comparison):} How do verifier-native FL techniques compare to a general-purpose LLM baseline?}

\begin{table}[t]    
    \caption{Fault localization performance across strategies (500 programs). Metrics as defined in Table~\ref{tab:rq1_all}.}
    \centering
    \small
    \begin{tabular}{lcccc}
        \toprule
        \textbf{Strat.} & \textbf{EX$_F$} & \textbf{EX$_M$} & \textbf{Top-1/3/5} & \textbf{F / E} \\
        \midrule
        CNTM & 0.109 & 0.155 & 43 / 71 / 78 & 84.00 / 1.40 \\
                
        LLM & 0.159 & 0.217 & 58 / 65 / 68 & 72.60 / 1.00 \\
        LLM\_ER & 0.264 & 0.303 & 37 / 48 / 50 & 50.80 / 0.00 \\
        LLM\_ER\_CN & 0.131 & 0.167 & 46 / 72 / 76 & 77.20 / 0.20 \\
        \bottomrule
    \end{tabular}
    \label{tab:rq4_all}
\end{table}
           
We compare CNTM with the LLM localization baseline and the LLM\_ER and LLM\_ER\_CN variants to contextualize verifier-native techniques within a broader landscape. The results are presented in \Cref{tab:rq4_all}, and show that LLM achieves higher Top-1 accuracy (58\%) than CNTM (43\%) (statistically significant), indicating stronger performance at placing the faulty line at the very top of the ranking. This behavior is further illustrated in Figure~\ref{fig:file-boxplot}, where the LLM variants exhibit a median closer to zero than CNTM, indicating that they more frequently rank the faulty line near the top. However, they also exhibit substantially longer tails, reflecting many cases in which the faulty line is ranked much lower or absent among the returned candidate lines.

In contrast, CNTM achieves better overall ranking quality as measured by EXAM. At file scope (however, not statistically significant), CNTM attains an EXAM of 0.109, compared to 0.159 for the LLM, suggesting lower average inspection effort across the full ranking. Furthermore, CNTM exhibits higher reliability, with a higher \emph{Found} rate (84.00\% vs.\ 72.60\%).

These results highlight complementary strengths between the approaches. LLM is effective at prioritizing the most likely fault location (Top-1), while CNTM provides more consistent rankings and broader coverage when the top prediction is insufficient. However, the LLM baseline is intentionally under-optimized, which likely underestimates the true potential of LLM-based approaches. We use a minimal setup (code-only input without prompt engineering, few-shot examples, or diagnostic signals such as verifier error messages) to provide a neutral reference point rather than a fully competitive configuration. To strengthen the comparison, we evaluated the two additional variants LLM\_ER and LLM\_ER\_CN.

LLM\_ER performs worse than the baseline, consistent with prior observations in Dafny \cite{mugnier2025laurel}. Error messages are often coarse-grained and typically indicate only the failing method, which, since the model is usually good at finding the method, adds noise rather than a useful localization signal. In practice, for this variant, the model frequently predicts the line of the failing postcondition, an unsurprising but uninformative answer, since that location cannot correspond to the root cause.

LLM\_ER\_CN also does not improve over either standalone method, suggesting that simple combinations are insufficient and more careful integration is required to realize any benefit.

A plausible explanation for CNTM's weaker Top-1 performance is that its heuristic scoring effectively identifies the relevant code region, but assigns equal suspicion to all lines in that block with an associated counterexample state, being incapable of finer-grained findings. In other words, CNTM is good at narrowing down where the fault lies, but less effective at deciding which specific line is most suspicious. This is precisely where LLMs may have an advantage, given their ability to reason over local code semantics.

\textbf{Advantages and Disadvantages.}
LLMs can use learned semantic patterns to achieve strong Top-1 performance, but are inherently non-deterministic and depend on external inference. In contrast, CNTM produces deterministic and reproducible rankings grounded in counterexample traces and operates entirely within the verification pipeline.

\begin{tcolorbox}[myboxstyle]
LLM achieves higher Top-1 accuracy, while CNTM provides more consistent ranking quality and reliability. These approaches should be viewed as complementary, and future work may explore hybrid methods combining counterexample information with LLM reasoning.
\end{tcolorbox}

\section{Threats to Validity}
\label{sec:threats}

\noindent\noindent
\emph{\textbf{Mutation realism.}} Our mutations are synthetic, generated by predefined operators. Real verification bugs may differ: they might involve complex logic errors, subtle contract misspecifications, or edge cases not well-modeled by simple mutations. However, mutation-based evaluation is standard practice in fault localization 
\cite{li2018mure}
and provides controlled ground truth unavailable in real-world settings. In addition, the mutation operators used in our dataset are drawn from existing work \cite{amaral2025mutdafny}, where several operators resulted from an analysis of bugfix evidence from Dafny projects
available on GitHub. 

\noindent\noindent
\emph{\textbf{Dataset characteristics.}} Since our dataset is drawn from DafnyBench, a collection of pedagogical and algorithmic programs, our results may not generalize, as industrial verification code may exhibit different patterns. 
However, DafnyBench is widely used in prior work and serves as a standard benchmark.

\noindent\noindent
\emph{\textbf{LLM comparison setup.}} The LLM baseline is included as a reference point to situate our verifier-native techniques within the emerging landscape of LLM-based fault localization approaches, not as an optimized competitor. Unlike classical test-based methods, LLMs require no test oracle and can be applied directly to source code. We deliberately avoid prompt engineering and few-shot examples for two reasons: (1) optimizing LLM performance is orthogonal to our research focus, and (2) a code-only input represents a lower bound on what an LLM can achieve without verifier integration. The reported results should be interpreted as indicative of the relative strengths of symbolic and learned approaches, and not as a ranking of which approach is superior. The further exploration of hybrid techniques is left to future work.

\section{Related Work}
\label{sec:related-work}

Traditional fault localization approaches mostly rely on test cases. SBFL~\cite{sbfl1, sbfl2} compares the execution frequency of program statements in failing tests as opposed to passing tests, to pinpoint those most likely to be the cause of the fault. This is measured, for each statement, by a \textit{suspiciousness score}, its value being higher the more a statement appears in failing tests and the less it appears in passing tests. In turn, a statement is more \textit{suspicious} the higher its suspiciousness score is. Several coefficients for computing this score have been proposed, e.g., Tarantula~\cite{tarantula1}, 
Ochiai~\cite{sbfl2}, and Barinel~\cite{barinel}.

MBFL~\cite{mbfl1}
extends SBFL by considering, in addition to execution frequency, how often mutating a statement affects the outcome of failing and passing tests. The more often mutating a statement changes the outcome of failing tests and the less it changes the outcome of passing tests, the more suspicious the statement is considered. The idea behind this is that these statements contribute to test failures and, therefore, to the faulty behavior.

AutoFix~\cite{AutoFix3}, for Eiffel, is the first Automated Program Repair (APR) tool to directly incorporate contracts into the fault localization process, along with tests automatically generated from those contracts. Instead of ranking suspicious statements, it ranks suspicious expressions that hold at certain locations, representing specific program states in which the program does not behave as expected. The goal of the fix generation phase is to attempt to avoid this state. Candidate fixes are generated from templates that account for the suspicious expression.

Maple~\cite{Maple} is an APR tool whose fault localization stage uses Hoare logic to determine the program state properties that hold at each point, and then identify violations of expected logical entailments regarding the specification. This approach has also been adapted for Dafny~\cite{verifixer-llms}. However, in both cases, it suffers from the limitation of only identifying the path containing the faulty location, rather than specific statements.

Sullivan et al. \cite{sullivan2017automated} studied test and mutation generation for Alloy. Although Alloy and Dafny operate under different paradigms (relational modeling versus imperative verification), their work aligns with ours in exploring the generation of diverse inputs from formal constraints and using mutation analysis to evaluate verification frameworks.
Zheng et al. propose FLACK~\cite{Zheng2021FLACK}, which localizes faults in Alloy models by analyzing counterexamples produced by bounded analysis and ranking model elements. Our work shares the high-level idea that counterexamples are first-class debugging artifacts, but differs in two core ways: (1) we target \emph{verified imperative programs} (Dafny) rather than relational models; (2) we localize \emph{source-code lines} and incorporate control-structure-aware ranking extracted from Dafny traces and instrumentation.

Beyer et al.~\cite{beyer2024fault} study fault localization using verification witnesses, i.e., artifacts that certify reachability properties and can be post-processed to identify likely fault sites. The witness perspective is complementary to ours, since it generalizes across verifiers and intermediate formats, whereas we exploit \emph{Dafny-specific} counterexamples to directly rank source lines.

\section{Conclusion}
\label{sec:conclusion}

This paper presents a study of automated fault localization for the Dafny verification-aware language. We evaluated state-based, counterexample-based, and LLM-based baselines across 500 mutants to better understand their relative strengths in identifying root causes of verification failures.

Our results show that counterexample-based localization is substantially more reliable than state-based approaches in this setting. SNAP exhibits a high \emph{Empty} output rate due to its reliance on a test suite, whereas counterexample-based techniques consistently produce actionable results, with \emph{Found} rates exceeding 84\%. Within this family, moving from raw traces (CNTB) to structured ranking (CNTS) yields a significant improvement, and multi-path aggregation (CNTM) further improves effectiveness, reaching a Top-1 accuracy of 43\%.

The comparison with the LLM baseline should be interpreted as contextual rather than competitive. The primary goal of this work is to study deterministic, verifier-native fault localization techniques, and not to develop or optimize LLM-based methods. Accordingly, the LLM is evaluated in a minimal configuration, serving as a reference point rather than a tuned baseline. In this setting, the LLM achieves higher Top-1 accuracy, while CNTM provides better overall ranking quality (lower EXAM) and higher reliability (\emph{Found} rate).

Overall, these results suggest that counterexample-based techniques provide a robust and consistent foundation for fault localization in verification-aware languages. At the same time, the observed strengths of LLMs in prioritization indicate potential benefits in combining the two approaches. Exploring hybrid methods that integrate counterexample information with LLM reasoning is a promising direction for future work.

\section{Data Availability Statement}

The full package required to reproduce the results is available in a pre-built, self-contained Docker image at \href{https://zenodo.org/records/19748378?token=eyJhbGciOiJIUzUxMiJ9.eyJpZCI6ImRkMGIxMmU0LWFlZmEtNGFhNS1iYWI3LTYwZjAxZWRkZDhiZiIsImRhdGEiOnt9LCJyYW5kb20iOiI2YjA1YWNmNTFlYzEzMzllOGY4MGJmOTMwY2UwN2EyNSJ9.-19PM43c4A5x5b7q1JapdZRoBQIjo5ZqLsyqalN4GrGV3fc_TCV4C7wCoH0kgumshEJTlO6pI7FgxaQJ3Ao-Tg}{Zenodo}.

\section*{Acknowledgments}

Isabel Amaral, João Pascoal Faria, and Alexandra Mendes were funded by National Funds through the FCT - Fundação para a Ciência e a Tecnologia, I.P. (Portuguese Foundation for Science
and Technology) within the project VeriFixer, with reference 2023.15557.PEX (DOI: 10.54499/2023.15557.PEX).

Álvaro F. Silva was co-financed by national funds through FCT – Fundação para a Ciência e a Tecnologia, I.P., under the support UID/50014/2025 (https://doi.org/10.54499/UID/50014/2025), and Fundação para a Ciência e a Tecnologia (Portuguese Foundation for Science and Technology) through the Carnegie Mellon Portugal Program under the fellowship reference PRT/BD/155045/2024.

\appendices
\section{Annex}
\begin{table}[ht]

\caption{Pairwise Wilcoxon signed-rank tests for file-level EXAM scores. Columns report compared methods (M.A, M.B), non-zero differences (NZ), statistic (W), $p$-value, rank-biserial effect size (R-bi), and significance (Sig., $\alpha=0.05$).}

    \centering \scriptsize
    \begin{tabular}{l|l|r|r|r|r|c}
        \hline
        \textbf{M.A} & \textbf{M.B} & \textbf{NZ} & \textbf{W} & \textbf{p} & \textbf{R-bi} & \textbf{Sig.} \\
        \hline
        RAND & CNTB & 476 & 48048 & 0.0037 & -0.154 & yes \\
        RAND & CNTS & 464 & 29196 & 1e-17 & 0.459 & yes \\
        RAND & CNTM & 467 & 24135 & 1e-25 & 0.558 & yes \\
        RAND & LLM & 468 & 44330 & 0.000317 & 0.192 & yes \\
        RAND & SNAP & 467 & 26946 & 2e-21 & -0.507 & yes \\
        CNTB & CNTS & 352 & 5284 & 2e-41 & 0.830 & yes \\
        CNTB & CNTM & 403 & 8116 & 4e-44 & 0.801 & yes \\
        CNTB & LLM & 452 & 29727 & 1e-14 & 0.419 & yes \\
        CNTB & SNAP & 442 & 32633 & 1e-9 & -0.333 & yes \\
        CNTS & CNTM & 288 & 11302 & 2e-11 & 0.457 & yes \\
        CNTS & LLM & 404 & 37547 & 0.153 & 0.082 & no \\
        CNTS & SNAP & 445 & 11546 & 1e-44 & -0.767 & yes \\
        CNTM & LLM & 338 & 25696 & 0.101 & -0.103 & no \\
        CNTM & SNAP & 450 & 9126 & 2e-51 & -0.820 & yes \\
        LLM & SNAP & 462 & 20552 & 2e-30 & -0.616 & yes \\
        \hline
    \end{tabular}

\label{tab:pairwise_wilcoxon_file}
\end{table}

\begin{table}[ht]

    \caption{Pairwise McNemar tests for file-level Top-1 localization success. Columns report methods (M.A, M.B), discordant pairs (Disc.), A/B correct only (A-O, B-O), $p$-value, paired odds ratio (OR), and significance (Sig., $\alpha=0.05$).}
    \label{tab:pairwise_mcnemar_top1_file}
    \centering\scriptsize
    \begin{tabular}{l|l|r|r|r|r|r|c}
        \hline
        \textbf{M.A} & \textbf{M.B} & \textbf{Disc.} & \textbf{A-O} & \textbf{B-O} & \textbf{p} & \textbf{OR} & \textbf{Sig.} \\
        \hline
        RAND & CNTB & 76 & 24 & 52 & 0.00176 & 0.467 & yes \\
        RAND & CNTS & 143 & 21 & 122 & 2e-18 & 0.176 & yes \\
        RAND & CNTM & 216 & 17 & 199 & 1e-40 & 0.088 & yes \\
        RAND & LLM & 270 & 7 & 263 & 2e-68 & 0.028 & yes \\
        RAND & SNAP & 56 & 25 & 31 & 0.504 & 0.810 & no \\
        CNTB & CNTS & 105 & 16 & 89 & 2e-13 & 0.184 & yes \\
        CNTB & CNTM & 186 & 16 & 170 & 1e-33 & 0.097 & yes \\
        CNTB & LLM & 276 & 24 & 252 & 4e-49 & 0.097 & yes \\
        CNTB & SNAP & 62 & 42 & 20 & 0.00715 & 2.073 & yes \\
        CNTS & CNTM & 111 & 15 & 96 & 1e-15 & 0.161 & yes \\
        CNTS & LLM & 265 & 55 & 210 & 2e-22 & 0.264 & yes \\
        CNTS & SNAP & 133 & 114 & 19 & 1e-17 & 5.872 & yes \\
        CNTM & LLM & 208 & 67 & 141 & 3e-7 & 0.477 & yes \\
        CNTM & SNAP & 204 & 190 & 14 & 1e-40 & 13.138 & yes \\
        LLM & SNAP & 268 & 259 & 9 & 7e-65 & 27.316 & yes \\
        \hline
    \end{tabular}

\end{table}

\bibliographystyle{ieeetr}
\bibliography{references}

\end{document}